# All-fiber highly efficient delivery of 2 kW laser over 2.45 km hollow-core fiber

Jing Shi[1,2,†], Binyu Rao[1,2,†], Zilun Chen[1,2,†,*], Zefeng Wang[1,2,†,*], Guangrong Sun[1,2], Zuyin Xu[4], Zhen Huang[1,2], Peng Li[3], Zihan Dong[3], Min Fu[1,2], Xin Tian[1,2], Baolai Yang[1,2], Jian Zhang[1,2], Zhiyue Zhou[1,2], Tianyu Li[1,2], Lei Zhang[3], Biao Shui[4], Chenxin Gao[1,2] and Jinbao Chen[1,2,*]

[1]College of Advanced Interdisciplinary Studies, National University of Defense Technology, Changsha, 410073, China;
[2]Nanhu Laser Laboratory, National University of Defense Technology, Changsha, 410073, China;
[3]State Key Laboratory of Optical Fiber and Cable Manufacture Technology, Yangtze Optical Fiber and Cable Joint Stock Limited Company (YOFC), Wuhan, 430074, China.
[4]Everfoton Technologies Corporation Limited, Wuhan, 430073, China.
[†]These authors contributed equally to this work.
*Corresponding authors: zilun2003@163.com. zefengwang_nudt@163.com. kdchenjinbao@aliyun.com.

## Abstract
Anti-resonant hollow-core fibers (AR-HCFs) have emerged as an important medium for high-power laser delivery due to their low optical nonlinearity and high damage threshold. However, current delivery systems of high-power laser based on AR-HCFs mainly rely on free-space optical components, which limits long-term stability in dynamic environments. Here, we report an all-fiber delivery of 2 kW laser with 85.3% transmission efficiency over 2.45 km, using a self-fabricated AR-HCF with a record low transmission loss of 0.175 dB/km at 1080 nm. This represents a nearly 500-fold improvement in the power-distance product compared to reported all-fiber AR-HCF-based laser transmission systems, achieving a record transmission distance for high-power laser delivery. Notably, we observed the phenomenon of stimulated Raman scattering amplified within the silica nested tubes in AR-HCF for the first time. By effectively suppressing the Raman noise from the laser source, we achieve an all-fiber laser delivery without stimulated Raman scattering of silica glass. This work marks a significant breakthrough in multi-kilometer and multi-kilowatt power delivery that is potentially useful for industrial manufacturing, nuclear decommissioning, laser drilling of oil, particle acceleration and so on.

## Introduction

High-power fiber lasers have found extensive applications in industry process, medical biotechnology, and scientific research[1-3], owing to their prominent advantages including excellent beam quality, high conversion efficiency, compact structure, and effective thermal management. In recent years, fiber lasers have continuously developed[4-6], and significantly expanded their application domains. Meanwhile, there is a growing demand for the flexible long-distance delivery of high-power lasers. For instance, flexible long-distance high-power laser delivery systems enable physical separation between laser sources and workpieces, improving factory layout flexibility[3]. In nuclear decommissioning applications, such systems facilitate remote cutting of radioactive structures, significantly reducing radiation exposure risks[7]. For oil and gas exploration, the technology enables safer and more precise subsurface laser drilling operations, providing a sustainable alternative to conventional rock fragmentation methods[8]. However, conventional solid-core fibers (SCFs) for laser delivery are limited by the material damage threshold and nonlinear effects, which restrict further increases in transmission power and length. For example, the ytterbium-doped fiber laser developed by Fujikura Ltd. demonstrates a transmission length of 20 m at 5 kW output power[5], which decreases to 3 m when the power is increased to 8 kW[6]. Anti-resonant hollow-core fibers (AR-HCFs) provide a novel approach for addressing the limitations of SCFs[9-15], which emerge as an enabling tool for the flexible long-distance delivery of high-power lasers.

AR-HCFs confine the propagation of light within an air-filled core through a micro-structured cladding[14]. This design minimizes the overlap between the light field and the surrounding silica glass to approximately $10^{-5}$, significantly reducing optical nonlinearity and enhancing the damage threshold[16,17]. Furthermore, with the development of AR-HCFs[18-22], the transmission loss of which has been comparable to that of SCFs. In C-band, AR-HCFs have achieved an exceptionally low loss below 0.1 dB/km, establishing a new record[23]. Notably, the laser operating in the 1 μm spectral band represents the highest power level of fiber laser output, wherein existing literature documents a transmission loss of 0.3 dB/km for AR-HCF[24]. These combined properties render AR-HCFs particularly advantageous for flexible long-distance delivery of high-power lasers. Notable demonstrations include the delivery of 1 kW continuous laser over 1 km with a loss of 0.74 dB/km[25], narrow-linewidth laser delivery of 2.2 kW over 100 m with a loss of 0.79 dB/km[26], and 3 kW multimode laser delivery over 100 m with a loss of 3.27 dB/km[27] based on AR-HCFs in the 1 μm spectral band, proving the potential of AR-HCFs for flexible long-distance delivery of high-power lasers.

However, existing high-power laser delivery systems based on AR-HCFs mainly rely on free-space optical components[26-30], which limits long-term stability in dynamic environments. In addition, high-power laser delivery introduces thermal lensing effects, which affects coupling efficiency in spatial coupling[27]. In terms of all-fiber laser delivery system, a delivery of 100 W single-frequency laser power over a 100 m AR-HCF has been reported in 2024[31]. However, the coupling method relies on fiber alignment through a five-axis adjustment stage, which requires precise operation. Moreover, the coupling position reaches a temperature of 84.6 °C at an output power of 100.3 W, preventing its application in kilowatt-class laser delivery.

In this work, we demonstrate an all-fiber delivery of 2 kW laser with 85.3% transmission efficiency over a 2.45 km AR-HCF designed and fabricated by ourselves. This represents a nearly 500-fold improvement in the power-distance product compared to reported all-fiber AR-HCF-based transmission systems[31], achieving a record transmission distance for high-power laser delivery. This achievement can be attributed to the following four key factors: Firstly, the AR-HCF achieves a record low transmission loss of 0.175 dB/km at 1080 nm; Secondly, a low-loss fusion splicing is developed between AR-HCFs and anti-reflection-coated SCFs, achieving a splicing loss less than 0.2 dB and a return loss less than -28.7 dB; Thirdly, we observed the phenomenon of stimulated Raman scattering (SRS) amplified within the silica nested tubes in AR-HCF for the first time. By effectively suppressing the Raman noise from the laser source by utilizing a chirped and tilted Bragg grating (CTFBG), we achieve an all-fiber laser delivery without SRS of silica glass. Lastly, an end cap, which is spliced with the output of AR-HCF, enables the plug-and-play functionality. The all-fiber integrated configuration constitutes a significant breakthrough, facilitating the transition of AR-HCF technology from the laboratory to practical applications.

## Results

### 1. Fiber design and fabrication.

The high-power laser source used in this work is a commercial laser source with a central wavelength of 1080 nm, utilizing a SCF of 20/250 μm (core/cladding diameter) as the output fiber, which supports two modes ($LP_{01}/LP_{11}$). In an all-fiber architecture, optimal coupling efficiency requires precise mode field matching between AR-HCFs and SCFs. Therefore, we designed an AR-HCF with a modal field matching with the 20/250 μm SCF. Furthermore, the designed AR-HCF could exhibit low transmission losses for both the fundamental ($LP_{01}$) and first-order ($LP_{11}$) modes to ensure efficient power delivery.

Currently, as is reported, an AR-HCF with a double-nested five-element structure has achieved a minimum transmission loss of 0.3 dB/km in the 1 μm spectral band[24]. Notably, another report demonstrates that increasing the number of nested tubes can effectively suppress cladding mode coupling, thereby significantly reducing transmission loss[32]. Basing on these studies, we designed a five-element triple-nested AR-HCF, where the transmitted light is strongly anti-coupling with cladding modes. The schematic diagram of the structure of the AR-HCF is shown in Fig. 1a, specifying the core diameter ($D_c$) and the diameters of the triple-nested tubes ($D$, $d$, and $d_n$). To achieve low transmission loss of $LP_{01}$ and $LP_{11}$ modes in the AR-HCF, we conducted systematic parameter exploration using systematic finite element analysis (FEA), as shown in Fig. 1b. It can be observed that when the parameter $D/D_c$ satisfies greater than 0.9, the transmission loss of the two modes stabilizes. Furthermore, the modal transmission loss exhibits a non-monotonic dependence on the scaling ratios between diameters of the adjacent tube: $d/D$ and $d_n/d$, where an optimal range exists. Across the studied ranges ($0.66 \leq d/D \leq 0.76$, $0.41 \leq d_n/d \leq 0.66$), the $LP_{01}$ and $LP_{11}$ modes demonstrate transmission losses below $10^{-3}$ dB/km and $10^{-2}$ dB/km, respectively. This occurs because undersized secondary/tertiary layers weaken anti-resonant confinement, causing substantial optical leakage and increased propagation loss. Conversely, oversized layers make adjacent nested tubes too close, inducing resonances that degrade transmission performance and consequently increase transmission loss. The blue regions in Fig. 1b show the low-loss regions with optimal parameters, providing a relatively large range for drawing fibers. The red box represents the parameters of the fabricated AR-HCF.

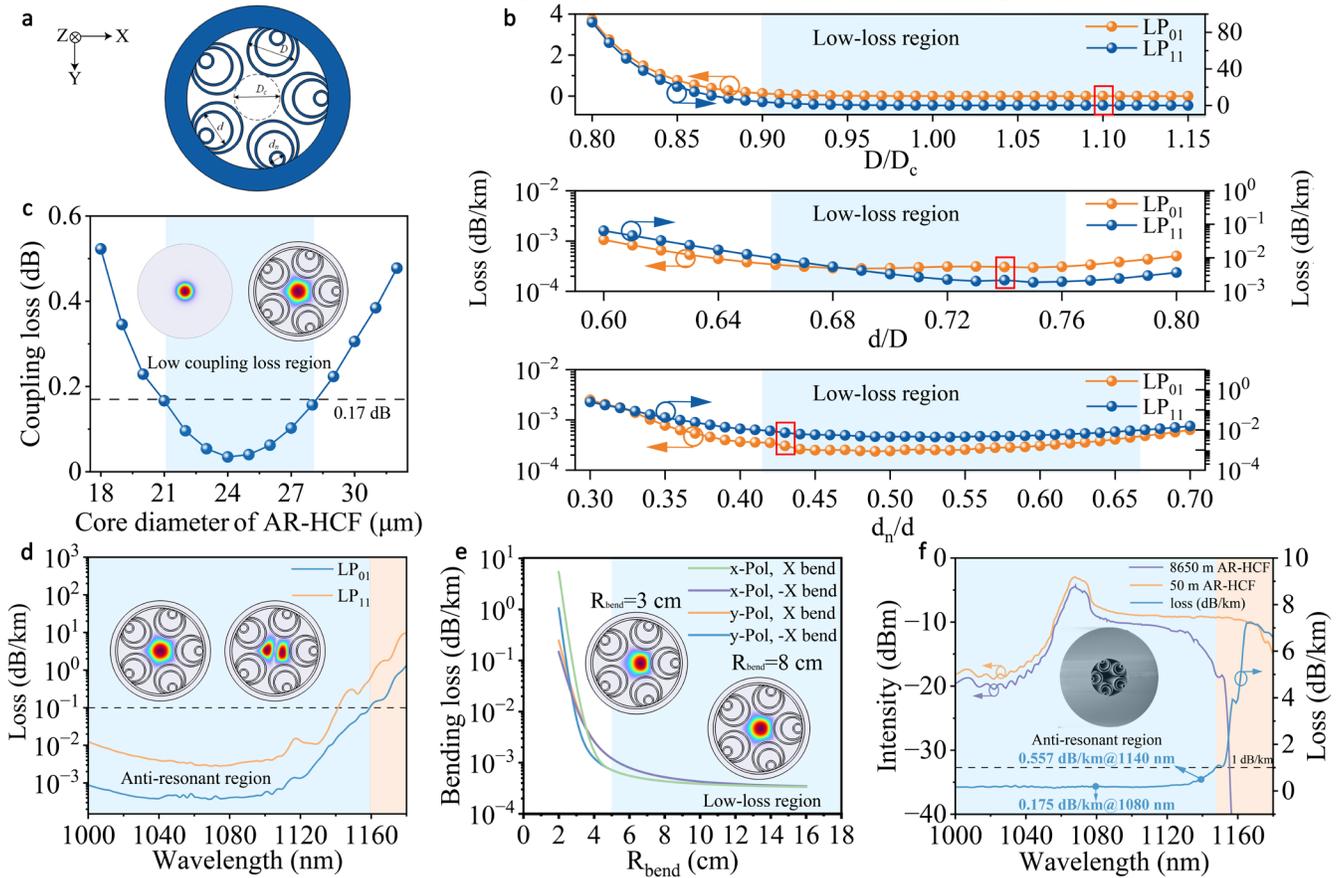

**Fig. 1: Characterization of the AR-HCF. a**, Cross section of the designed AR-HCF. **b**, Simulated loss of the $LP_{01}$ (orange line), $LP_{11}$ (blue line) core modes as a function of $D/D_c$, $d/D$, and $d_n/d$, respectively. **c**, Relationship between coupling loss and core diameter of AR-HCF. Insets show fundamental mode-field profiles of SCF and AR-HCF. The blue regions show low-loss regions. The red box represents the actual parameters of the fabricated AR-HCF. **d**, Simulated loss of the $LP_{01}$ (blue line), $LP_{11}$ (orange line) core modes as a function of wavelength. The insets show mode-field profiles of $LP_{01}$ and $LP_{11}$ core modes of the AR-HCF. **e**, The bending loss of AR-HCF as a function of bending radius. The insets show fundamental mode-field profiles of AR-HCF at bend radii of 3 cm and 8 cm. **f**, Measured transmission spectra and the evaluated cut-back loss of the 8650 m long and 50 m long AR-HCF. The inset shows SEM image of the uncoated fiber cross-section.

Under high-power laser transmission, mode field matching is critical, as mode field mismatches can induce significant coupling loss, localized heating at fusion points, and even potential system failure. Figure 1c shows the coupling loss induced by mode field mismatch between AR-HCF with a core diameter of 18-32 μm and the SCF with a core diameter of 20 μm. The fundamental mode profiles of the SCF and the AR-HCF were simulated, as shown in the insets of Fig. 1c. The coupling loss at the butt-coupled interface can be calculated as[33]

$$\alpha = -10\log_{10}(\frac{|\int E_i^* E_t dS|^2}{\int E_i^* E_i dS \int E_t^* E_t dS}) \quad (1)$$

Here, $E_i$ and $E_t$ represent the transverse electric fields of the fundamental mode in the SCF and AR-HCF, respectively. Numerical simulations reveal that when the $D_c$ is within the 21-28 μm range, the coupling loss is below 0.17 dB, corresponding to a theoretical coupling efficiency exceeds 96%. Theoretical analysis indicates that increasing the core diameter of AR-HCFs can effectively reduce the coupling efficiency between the mode field and the nested tubes, thereby significantly suppressing mode leakage. Therefore, this study employs a core diameter of 28 μm. Table 1 shows the parameters of the fabricated AR-HCF. The average thickness of the nested tubes is 1.3 μm, which places the target wavelength of 1080 nm in the anti-resonance region of the second window[34].

**Table 1 Parameters of fabricated AR-HCF**

| $D_c$ (μm) | $D$ (μm) | $d$ (μm) | $d_n$ (μm) | $D/D_c$ | $d/D$ | $d_n/d$ |
|---|---|---|---|---|---|---|
| 28 | 31 | 23 | 10 | 1.1 | 0.74 | 0.44 |

The simulation result of the $LP_{01}$ and $LP_{11}$ modes losses is shown in Fig. 1d under fabricated parameters. The entire area in Fig. 1d is divided into anti-resonant region (blue) where the $LP_{01}$ mode maintains losses below 0.1 dB/km, and resonant region (orange). At the wavelength of 1080 nm, the $LP_{01}$ mode exhibits confinement losses below $10^{-3}$ dB/km, while the $LP_{11}$ mode remains under $10^{-2}$ dB/km. The insets in Fig. 1d show mode-field profiles of $LP_{01}$ and $LP_{11}$ core modes of the AR-HCF. Figure 1e presents the simulated bending losses for both x-polarization and y-polarization in the ±X-direction. The results indicate that the fiber exhibits excellent bending resistance with values below $10^{-3}$ dB/km for bending radius exceeding 5 cm. The insets in Fig. 1e show simulated mode profiles of AR-HCF at bend radii of 3 cm and 8 cm. Figure 1f shows the cutback loss, with measurements conducted over lengths ranging from 8650 m to 50 m using a supercontinuum source and an optical spectrum analyzer (OSA) (see Methods). The anti-resonant region (blue) is defined where the transmission loss below 1 dB/km. The bending radius of AR-HCF is 8 cm. The results demonstrate a loss of 0.175 dB/km at 1080 nm, which is currently the record low-loss value in this spectral band. At 1140 nm, the loss of AR-HCF is 0.557 dB/km. The loss of the fabricated AR-HCF is higher than expected. This is due to optical leakage caused by micro-bending, surface scattering and minor deformations of partial nested tubes[11,27].

## 2. Low-loss fusion splicing.

Due to the different materials at the interface between SCFs and AR-HCFs (silica/air), back-reflected light induced by Fresnel reflection accounts for approximately 4% of the input power. Here, the anti-reflection coating method[35] was adopted to decrease the energy loss caused by Fresnel reflection. The end faces of the SCFs were coated with anti-reflection coatings, achieving 0.4% reflectivity at 1080 nm. The AR-HCF and the anti-reflection-coated SCF were spliced using a commercial fiber splicer. Figure 2 presents the experimental setup. In this setup, a 1060 nm narrow-linewidth fiber laser operating at an output power of 10 mW was utilized as the light source, and back-reflected light was monitored through an optical circulator. The transmission efficiency is calculated as the ratio between the input power ($P_{in}$) and the output power ($P_{out}$) of the AR-HCF. The fusion splicing parameters were adjusted in real-time based on the results of transmission efficiency. The anti-reflection coating integrity can be evaluated through comparative analysis of $P_{back}$ before and after splicing processes. A significant increase in $P_{back}$ indicated compromised coating performance.

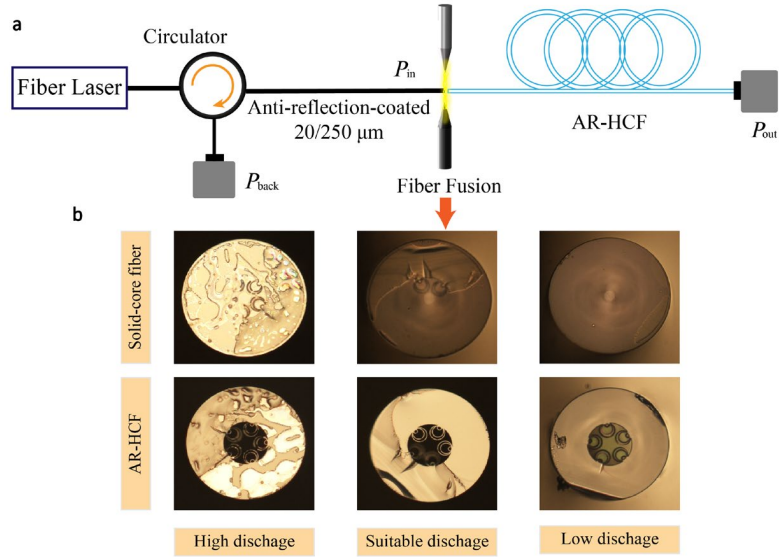

**Fig.2: Fiber fusion splicing of the anti-reflection-coated SCFs and the AR-HCF. a**, The experimental setup of fusion splicing. **b**, Cross-section views of post-splicing of SCFs and an AR-HCFs at relatively high, suitable, and low discharge parameters.

Experimental results demonstrate that discharge power and duration significantly influence splicing performance. Figure 2b provides a microstructural characterization of fusion-spliced fiber interfaces at relatively high, suitable, and low discharge parameters. Excessive discharge parameters induce thermal degradation of the anti-reflection coatings and collapse of the AR-HCF microstructure. Conversely, when discharge parameters are insufficient, the mechanical bonding at the fusion point is weak and easier to break, with no observable fusion traces remaining on the surface of the SCF. Optimal discharge energy and duration can achieve a balance. The anti-reflection coatings and AR-HCF capillaries remain intact, while robust mechanical bonding is established. Notably, the distinct imprint of the nested tubes on the SCF surface confirms precise geometric matching at the interface, with no observable coating delamination or thermal deformation. Under suitable parameters, fusion splicing losses are maintained below 0.2 dB, with return losses below −28.7 dB after several fusion tests. The coupling losses align closely with simulated predictions, validating the fidelity of the numerical model.

## 3.  Experimental system.

Figure 3 shows the experimental for high-power laser delivery. The laser source emitted a continuous wave of 1080 nm. A CTFBG was added after the laser source to suppress the Raman noise. The output laser was delivered via an anti-reflection-coated SCF of 20/250 μm, and was subsequently fusion-spliced with the AR-HCF using the optimized parameters. The fusion point was positioned within a glass tube for stability and protection. The AR-HCF was coiled on a 32-cm diameter drum. In practical applications, the output terminal of AR-HCF is prone to contamination from dust, pollutants, or environmental moisture. To address this, we fused a coated end cap at its output end for protection, enabling the plug-and-play functionality. The practical maps of the fusion point and the end cap are shown in the insets of Fig. 3.

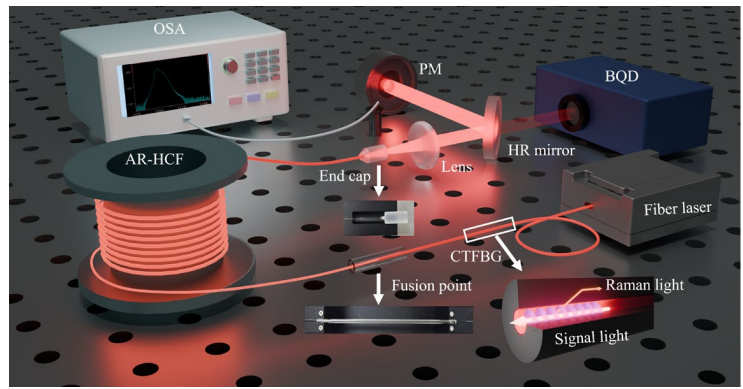

**Fig. 3: The experimental step of high-power laser delivery.** OSA, optical spectrum analyzer; PM, power monitor; BQD, beam quality detector; HR mirror, high reflection mirror; CTFBG, chirped and tilted Bragg grating. The insets show the practicality maps of fusion point, end cap of AR-HCF and the concept map of CTFBG.

## 4. Observation of SRS in AR-HCF.

Preliminary investigations were carried out utilizing a 200 m AR-HCF for high-power laser delivery, with no CTFBG employed in the setup. At an input power of 2400 W, spectral monitoring reveals characteristic peaks at around 1140 nm, which is considered as the Stokes light generated by SRS in silica glass material, as shown in Fig. 4a. To determine whether SRS in AR-HCF originated from the laser source, a CTFBG was spliced after the laser source, whose Raman suppression ratio exceeding 20 dB. In contrast, the control group was spliced with the same length SCF of 20/250 μm with the CTFBG. Figure 4a shows the output spectra of a 200 m AR-HCF at 2400 W input power, comparing cases with and without the CTFBG. The SRS component disappears when the CTFBG was inserted. This confirms that internal noise from the laser source is a key factor for inducing Raman scattering in AR-HCF. With the fusion point kept, distinct spectra were compared with transmission AR-HCF lengths of 2 m and 200 m under input powers of 2259 W, 2315 W, and 2400 W, respectively, as shown in Fig. 4b. By integrating the spectra, we determined that at 2259 W input power, the SRS intensity after a 200 m AR-HCF transmission is 2.98 times higher than that after a 2 m transmission, confirming nonlinear accumulation effects. It is worth noting that under input powers of 2315 W and 2400 W, the Raman amplification of 200 m AR-HCF relative to 2 m AR-HCF decreased to 1.33 and 1.28 times, respectively. This reversal phenomenon is mainly due to the differential attenuation characteristics in AR-HCF: when the signal light experiences a transmission loss of 0.175 dB/km, the attenuation of the Raman component is higher with a transmission loss of 0.557 dB/km. Since the loss of AR-HCF to Raman light is higher than that of signal light, Raman growth is slower in longer fibers as the power increases. Based on the scaling ratio of SRS in the experiments, we can calculate the effective Raman gain coefficient of AR-HCF as 2.4 $km^{-1} \cdot kW^{-1}$ (see Methods for the detailed define and calculation).

Figure 4d-f show the signal and Raman beam profiles, and their corresponding spectra (see Methods for detailed measurement). At an input power of 1005 W, the beam profile of signal light predominantly consists of core light, as shown in Fig. 4d. When the input power is 2259 W, the signal light and Raman light simultaneously exist and exhibit a mixed mode within both the fiber core and the nested tubes, characterized by a more pronounced intensity within the core (Fig. 4e). Figure 4f shows the spectrum and beam profile of only Raman light after filtering most of the signal light at an input power of 2259 W. It is discernible that Raman light exists in both core and nested tubes.

The experimental results demonstrate that SRS originates in Raman noise stemming from the laser source, and is then amplified in the nested tube structure. Furthermore, fiber offset during splicing results can increase overlap between the optical field and nested tubes of AR-HCF, which in turn enhances the Raman gain. We simulated and calculated the effective Raman coefficients of AR-HCF varying with fiber offset using FEA, and the results are shown in Fig. 4c. When the input laser is coupled into the fiber core without any deviation, the effective Raman coefficient is $2.8\times10^{-3}$ km$^{-1}\cdot$kW$^{-1}$. At a fiber offset of around 3 µm, the effective Raman coefficient increases sharply to 2.55 km$^{-1}\cdot$kW$^{-1}$, which is close to experimental results. This indicates a deviation of approximately 3 µm between the SCF and the AR-HCF at the fusion point. Furthermore, the effective Raman coefficient of the AR-HCF has a change due to various factors, including the precision of splicing alignment, the thickness of the nest tubes, the core diameter of AR-HCF and so on.

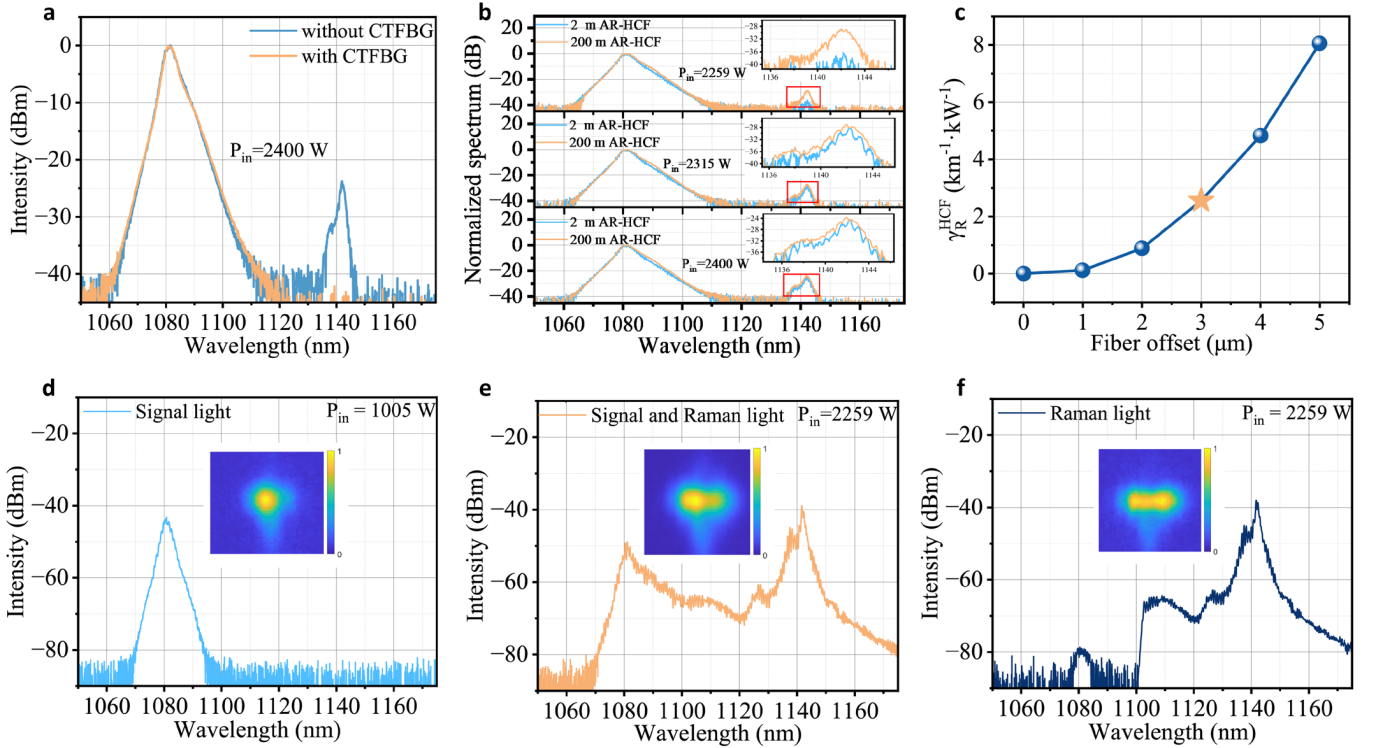

**Fig.4: Experiment results of SRS in AR-HCF. a**, Output spectrum of 200 m AR-HCF with CTFBG (orange line) and without CTFBG (blue line) at 2400 W input power. **b**, The normalized output spectrum of 2 m (blue line) and 200 m (orange line) AR-HCF at 2259 W, 2315 W and 2400 W input power at the same fusion point between SCF and AR-HCF. The red boxes are enlarged areas. **c**, Simulated the relationship between Raman coefficients and fusion offset. Transmission spectrum and beam profiles of **d**, signal light at 1005 W input power, **e**, signal and Raman light and **f**, Raman light at 2259 W input power over a 2 m AR-HCF.

## 5. High-power delivery in AR-HCFs.

Drawing from the preceding analysis of SRS, we conducted a laser delivery by utilizing a CTFBG to mitigate the intrinsic Raman noise of laser source, as illustrated in Fig. 4. Figure 5 presents the high-power laser delivery results over 1 km and 2.45 km. The power delivery performances of 1 km and 2.45 km AR-HCF systems with and without end caps are shown in Fig. 5a. The left axes show the transmission efficiency and the corresponding output power vary with the input power of AR-HCF. At a maximum input power of 2400 W, the 1 km AR-HCF delivers 2160 W with a transmission efficiency of 89.9%, which reduces to 2022 W with a transmission efficiency of 84.2% when an end cap is spliced. For a 2.45 km AR-HCF, the maximum output power is 2050 W with a transmission efficiency of 85.3%, while the maximum output power after the end cap is 1960 W with a transmission efficiency of 81.6%. Based on the power delivery results of 1 km and 2.45 km, the transmission loss of AR-HCF can be calculated as 0.172 dB/km. The measured transmission loss is marginally lower than the cutting loss (0.175 dB/km), primarily due to the reduced bending loss from the increased coiling diameter employed in our experimental setup. The coupling loss is obtained by subtracting the measured propagation loss of 0.172 dB/ km. At the maximum input power, the coupling loss falls within the range of 0.27-0.29 dB, which has a marginal increase in comparison to the results obtained during low-power testing (0.2 dB).

Figure 5b shows the output spectra of the source, 1 km AR-HCF and 2.45 km AR-HCF at the maximum input power. The output spectra of the AR-HCF exhibit excellent consistency with the laser source spectrum, with no observable SRS nonlinear effects. The minor spectral redshift can be attributed to the Raman response of the atmospheric air within the core[25]. Figure 5c presents a 2-hour power test of the end cap output structure with a 1 km AR-HCF, showing a power fluctuation of 2.3% at the maximum input power. Meanwhile, Fig. 5c displays the thermal image of the coiling fiber and glass tube with a fixed fusion point. The highest temperature observed in the glass tube is 58.59 °C. The average temperature of the coiling fiber is 22.19 °C, but there are individual hotspots with temperatures of 58.8 °C. At maximum input power (2400 W), the laser source exhibits a beam quality factor (M²) of 1.26. Corresponding measurements of the AR-HCF output (Fig. 5d) demonstrate excellent beam quality preservation, with M² values of 1.30 (1 km AR-HCF) and 1.29 (2.45 km AR-HCF) under identical power conditions. The near-field diffraction patterns of 1 km and 2.45 km AR-HCFs were systematically characterized under varying power levels, with measurements conducted at input powers of 1018 W, 1611 W, and 2400 W, as illustrated in Fig. 5e. The results demonstrate that as fiber length and power increases, the beam profiles consistently maintain a well-defined Gaussian distribution without significant degradation. This result demonstrates the capability of AR-HCFs to maintain beam quality over kilometer-scale distances at multi-kilowatt power levels, confirming their suitability for industrial and scientific high-power long-distance laser delivery systems.

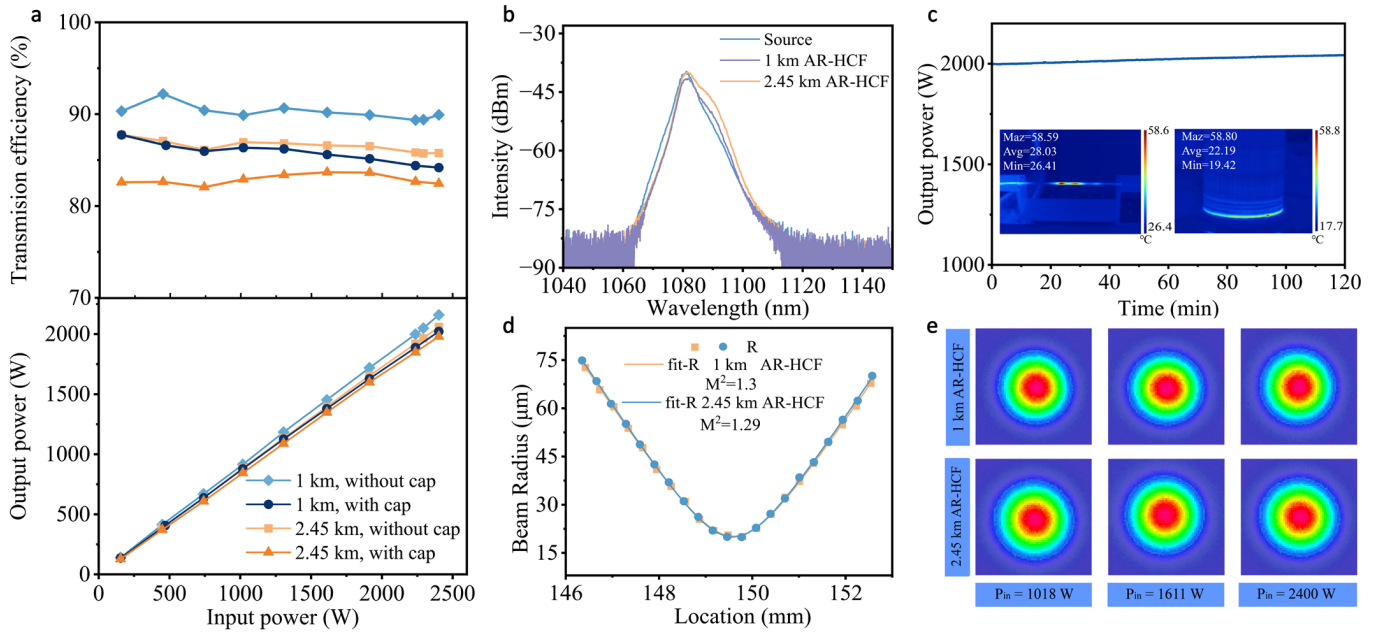

**Fig. 5: High-power laser delivery results. a,** Power delivery results in 1 km and 2.45 km. **b,** Input and output spectra of the 1 km and 2.45 km AR-HCF at maximum input power; **c,** 2-hour power test result of 1 km AR-HCF end cap output at the maximum input power. The inset shows the thermal image of the fusion point and fiber optic disc at maximum input power. **d,** Measurement of beam quality factor of 1 km and 2.45 km AR-HCF at the maximum input power; **e,** Beam profile diagram of AR-HCF at different input powers of 1018 W, 1611 W and 2400 W at different fiber lengths of 1 km and 2.45 km AR-HCF.

**Discussion and Conclusion**

In this study, we report an integrated all-fiber laser delivery system based on AR-HCFs, which provides a robust solution to spatial coupling instability in high-power applications. Our self-fabricated AR-HCF features an optimized five-element triple-nested architecture that achieves an ultralow transmission loss of 0.175 dB/km at 1080 nm, which represents the lowest reported value for this spectral band to date. Moreover, we developed an optimized fusion splicing technique for AR-HCF and anti-reflection-coated SCFs, achieving a low fusion splicing loss (<0.2 dB) and a return loss (<-28.7 dB). Furthermore, the fusion point between anti-reflection-coated SCF and AR-HCF was fixed with glass tubes, and the output end was spliced with an end cap for protection. Consequently, a stable all-fiber high-power of 2 kW delivery system was achieved over a 2.45 km AR-HCF with a transmission efficiency of 85.3% through the above technologies. This is of great value for facilitating the transition of AR-HCF technology from the laboratory to practical applications.

Notably, the SRS amplified within the silica nested tubes in AR–HCF was observed for the first time in our experiments, which was caused by the amplification of the Raman noise stemming from laser source. This nonlinear effect is considered as a limiting factor for achieving higher power and longer transmission distance. Two suppression strategies are proposed: Firstly, Raman noise from the laser source can be effectively suppressed through a CTFBG, which has been validated in our experiments; Secondly, the AR-HCF can be designed to reduce the Raman gain. Simulations reveal significant correlations between structural parameters (fiber offset, nested tube thickness, and core diameter) and effective Raman gain coefficients. In all-fiber architecture, as the fiber offset increases, the effective Raman gain coefficient also increases. The thickness of nested tubes exhibits a nonlinear response characteristic with respect to the effective Raman gain coefficients. When the nested tube thickness deviates the signal wavelength from the resonance condition, the overlap integral between the optical field and the nested tube decreases, thereby inducing an optimal gain coefficient. The fiber core diameter is negatively correlated with the effective Raman gain coefficient, so the core size should be maximized in design under the condition of ensuring fundamental mode transmission.

To investigate the scalability at higher power levels and longer transmission distances, we conducted numerical simulations accounting for two primary limiting factors: SRS amplified within the silica nested tubes and a 3 μm beam offset at the input due to imperfect alignment between the SCF and AR-HCF. Based on the SRS threshold criterion[36] (see Methods), Fig. 6 presents the achievable transmission lengths vary with output power for AR-HCFs with different core diameters. The simulations reveal that at 2 kW output power, AR-HCFs with core diameters of 25 μm and 30 μm achieve transmission distances of approximately 2.8 km and 6.82 km respectively, while for 10 kW operation, only fibers with core diameters exceeding 35 μm can maintain transmission beyond 1 km due to SRS limitations, demonstrating the critical role of core diameter scaling of AR-HCF in high-power fiber transmission systems.

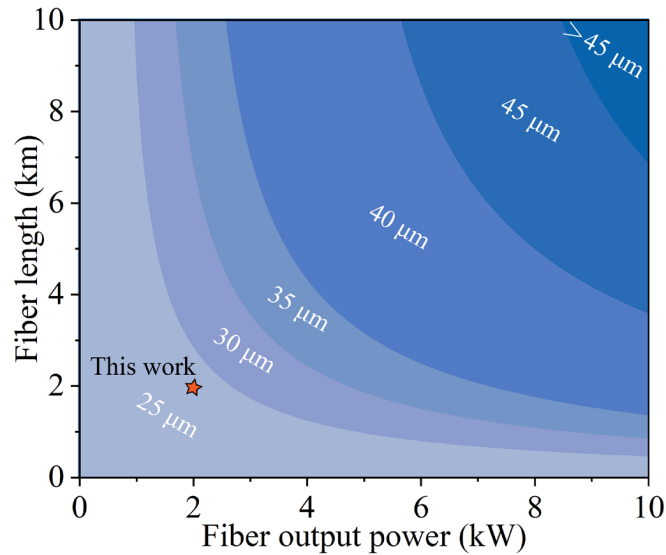

**Fig. 6: The scalability of continuous power transmission of AR-HCF in 1 μm spectral band, which is limited by SRS amplified within the silica nested tubes.** The spectral profile was determined based on the standard approximation of the SRS critical power (see Methods). The white text in the figure corresponds to the core diameter of AR-HCF. The orange pentagram represents the transmission power and transmission length corresponding to this work.

Looking forward, the all-fiber integrated laser delivery system based on AR-HCFs not only exhibits significant application potential in fields requiring high-power long-distance laser delivery, such as industrial manufacturing[3], nuclear decommissioning[7] and laser drilling of oil[8], but also has substantial value in other fields. Firstly, the long-distance transmission limitations of single-frequency lasers imposed by stimulated Brillouin scattering (SBS)[37] can be addressed through the AR-HCF due to its low nonlinearity and transmission loss. Secondly, long-distance AR-HCFs enable novel particle acceleration through radiation pressure, building upon experimentally demonstrated particle levitation and guidance in hollow-core fibers[38]. Thirdly, a low-loss all-fiber gas cell is constructed in this work, enabling high-sensitivity gas detection[39]. Moreover, our results may inspire other applications of AR-HCFs, such as fiber communications, distributed sensing, and quantum optics [40-42].

## Methods

### 1. AR-HCF loss measurement.

The loss of AR-HCF in this study was measured using the cutback technique. The test source was a supercontinuum source, which was connected to a PM-980 single-mode fiber as the launch fiber. The light source was coupled to an 8650 m AR-HCF featuring a bending diameter of 16 cm by employing the alignment function of the fusion splicer. Then the AR-HCF output was connected to the OSA using a bare fiber adapter. Firstly, the output spectrum was recorded using the 8650 m AR-HCF. The fiber was then cut to 50 m while maintaining identical emission conditions, and the output spectrum was measured again. The transmission loss was measured by comparing output spectra before and after fiber length reduction. Using long fibers for measurement can obtain more accurate loss values.

### 2. Signal and Raman light beam profile measurement.

The beam profiles of both the signal light and the Raman light were measured using a charge-coupled device (CCD). The output light from the AR-HCF was first collimated using a 100 mm focal-length plano-convex lens. Subsequently, a beam-splitting system composed of two high-reflectivity mirrors (HR mirrors) directed the majority of the beam energy served as a power attenuation mechanism, while the transmitted component was utilized for detection. The transmitted light was then focused by a 400 mm focal-length lens, with the CCD precisely positioned at the focal plane to capture the near-field diffraction pattern of the AR-HCF output. To accommodate dynamic power adjustment requirements, a tunable optical attenuator was incorporated upstream of the CCD to optimize the dynamic range of light intensity.

### 3. Calculated method of Raman gain coefficient of AR-HCF.

#### a. Calculation based on experimental data.

We can calculate the Raman gain coefficient of AR-HCF using the coupled wave equation:

$$\frac{dI_R}{dz} = g_R I_S I_R - \alpha_R I_R \qquad (2)$$

$$\frac{dI_S}{dz} = -\frac{\omega_S}{\omega_R} g_R I_S I_R - \alpha_S I_S \qquad (3)$$

When $I_R \ll I_S$, the transmission loss of fiber ($\alpha_R, \alpha_S$) can be ignored. Therefore,

$$G_{HCF} = \exp(\gamma_R^{HCF} P_S L_{eff}) \qquad (4)$$

$$L_{eff} = \frac{1 - e^{-\alpha_s L}}{\alpha_s} \qquad (5)$$

$$\gamma_R = \frac{g_R}{A_{eff}} \qquad (6)$$

$$G_{HCF} = \exp(\gamma_R^{HCF} PL - \alpha L) \qquad (7)$$

Here, $I_R$ and $I_S$ denote the intensities of the Raman and signal light, respectively, while their corresponding angular frequencies are represented by $\omega_R$ and $\omega_S$. The Raman gain characteristics are described by two coefficients: the Raman gain coefficient $g_R$ and the effective Raman gain coefficient $\gamma_R$. $\alpha_S$ and $\alpha_R$ represent the transmission loss of signal light and Raman light. $L_{eff}$ and $L$ represent the effective interaction length and transmission length, respectively. $A_{eff}$ is the effective mode field area. Experimental results demonstrate a total Raman gain $G_{HCF} \approx 2.98$ at a signal power ($P_S$) of 2259 W. With a signal attenuation $\alpha_S$ = 0.175 dB/km and $L_{eff} \approx L = 0.2$ km, we derive an effective Raman gain coefficient $\gamma_R^{HCF} \approx 2.4 \text{ km}^{-1} \cdot \text{kW}^{-1}$.

#### b. Calculation based on simulation results

Furthermore, the Raman gain coefficient of AR-HCF can be determined using the Raman gain coefficient of the solid core fiber as a reference.

$$\gamma_R^{HCF} = \eta \frac{A_{eff}}{A_{tubes}} \gamma_R^{SCF} \qquad (8)$$

For the SCF of 20/250 μm, its mode field area is approximately 284 μm² ($A_{eff}$). The Raman gain coefficient of SCF[36] $g_R$ is $10^{-13}$ km·kW⁻¹. Based on FEA, the proportion of the mode field within the nested tube ($\eta$) is determined. Here, due to that SRS amplified within the silica nested tubes in AR-HCF, the area of the nested tubes is used instead of the mode field area of Raman light in AR-HCF. Subsequently, $\gamma_R^{HCF}$ is calculated. Based on analysis.

According to the definition of the SRS critical power - which refers to the state where the power at the signal wavelength and Raman wavelength reach equilibrium - its expression can be approximated as[36]:

$$P_{\text{cr}} = \frac{16 A_{\text{eff}}}{g_R L_{\text{eff}}} = \frac{16}{\gamma_R L_{\text{eff}}} \quad (9)$$

According to the simulation calculation of $\gamma_R$, the maximum transmission power and length can be calculated, as shown in Fig. 6.

**Table 2 Calculation parameters related to SRS**

| SCF | | | AR-HCF | | |
|---|---|---|---|---|---|
| $g_R$ (km·kW$^{-1}$) | $A_{\text{eff}}$ (μm$^2$) | $\gamma_R^{\text{SCF}}$ (km$^{-1}$·kW$^{-1}$) | $\alpha_S$ (dB/km) | L (km) | $A_{\text{tubes}}$ (μm$^2$) |
| $1 \times 10^{-13}$ | 284 | 352 | 0.175 | 0.2 | $10^3$ |

## Data Availability
The data underlying the results presented in this paper are not publicly available at this time but may be obtained from the authors upon reasonable request.

## Acknowledgments
We gratefully acknowledge support from the Science and Technology Innovation Program of Hunan Province (2021RC4027). We would like to thank Yangmei Sun, Xiaoxi Liu, and Lianzhou Jiang for the sincere help and support on the fabrication and test.

## Author Contributions
Zilun Chen, Zefeng Wang and Jinbao Chen conceived the idea of the construction and methods for achieving all-fiber 2 kW laser delivery. Jing Shi and Guangrong Sun simulated the theoretical results. Zuyin Xu, Peng Li, Zihan Dong, Biao Shui, and Lei Zhang fabricated the AR-HCF. Jing Shi, Zilun Chen, Binyu Rao and Baolai Yang assembled the setup and performed the measurements. Jing Shi, Zilun Chen, Min Fu, Xin Tian, Jian Zhang and Tianyu Li performed the power delivery, loss and M$^2$ measurements. Zhen Huang assisted in measuring the output beam profile. Binyu Rao and Zhiyue Zhou contributed to the interpretation of the experimental results. Jing Shi, Binyu Rao, Guangrong Sun, Chenxin Gao and Zefeng Wang prepared the manuscript. Zefeng Wang supervised this work and led the scientific collaboration.

## Competing interests
The authors declare no conflicts of interest.